\documentclass{svmult}

\usepackage{makeidx}         
\usepackage{graphicx}        
\usepackage{multicol}        
\usepackage[bottom]{footmisc}
\usepackage{amsmath}
\usepackage{subfigure}
\usepackage{tabularx}
\usepackage{booktabs}

\makeindex             

\begin{document}

\title*{Experimental study of pedestrian flow through a T-junction}
\author{J Zhang\inst{1}\and
W Klingsch\inst{1}\and A Schadschneider\inst{2}\and A
Seyfried\inst{3,4}}
\institute{Institute for Building Material Technology and Fire
Safety Science, Wuppertal University, Pauluskirchstrasse 11, 42285
Wuppertal, Germany \texttt{ jun.zhang@uni-wuppertal.de,
klingsch@uni-wuppertal.de} \and Institut f\"ur Theoretische Physik,
Universit\"at zu K\"oln, 50937 K\"oln, Germany
\texttt{as@thp.uni-koeln.de} \and {Computer Simulation for Fire
Safety and Pedestrian Traffic, Wuppertal University,
Pauluskirchstrasse 11, 42285 Wuppertal, Germany}\and {J\"ulich
Supercomputing Centre, Forschungszentrum J\"ulich GmbH, 52425
J\"ulich, Germany} \texttt{seyfried@uni-wuppertal.de}}
%
%
\maketitle

\begin{abstract}
In this study, series of experiments under laboratory conditions
were carried out to investigate pedestrian flow through a
T-junction, i.e., two branches merging into the main stream. The
whole duration of the experiments was recorded by video cameras and
the trajectories of each pedestrian were extracted using the
software {\em Petrack} from these videos. The Voronoi method is used
to resolve the fine structure of the fundamental diagram and spatial
dependence of the measured quantities from trajectories. In our
study, only the data in the stationary state are used by analyzing
the time series of density and velocity. The density, velocity and
specific flow profiles are obtained by refining the size of the
measurement area (here $10~cm \times 10~cm$ are adopted). With such
a high resolution, the spatial distribution of density, velocity and
specific flow can be obtained separately and the regions with higher
value can be observed intuitively. Finally, the fundamental diagrams
of T-junction flow is compared in three different locations. It is
shown that the fundamental diagrams of the two branches match well.
However, the velocities in front of the merging are significantly
lower than that in the main stream at the same densities. After the
merging, the specific flow increases with the density $\rho$ till
$2.5~m^{-2}$. While in the branches, the specific flow is almost
independent of the density between $\rho =1.5~m^{-2}$ and
$3.5~m^{-2}$.
\end{abstract}

\section{Introduction}

In recent years, research on pedestrian and traffic flow became
popular and attracted a lot of attention
\cite{Appert-Rolland2009,Bandini2010,Klingsch2010,Schadschneider2009,Schadschneider2009c}.
A great deal of models has been proposed to simulate pedestrian
movement in different situations \cite{Helbing2000, Song2006}. Most
of these models are able to reproduce crowd phenomena such as lane
formation qualitatively. On the other hand, some field studies and
laboratory experiments are also carried out to obtain empirical data
and improve models.

Unfortunately, the empirical database is insufficient leading to
discrepancies among different handbooks. The fundamental diagram
describing the empirical relation between density ($\rho$), velocity
($v$) and flow ($J$) (or specific flow ($J_s$)) is the key
characteristics of pedestrian dynamics used to support the planning
and design of facilities. However, there is considerable
disagreement among the empirical data from different studies and
handbooks even for the same type of facility or movement. The
maximum flow as well as the density where the maximum flow appear
has a broad range \cite{Schadschneider2009a,Seyfried2009}. These
discrepancies cause big inconvenience in design and assessment.
Further research is necessary to study the reasons for these
differences. Carrying out laboratory experiments is a good way to do
this, since secondary factors can be removed or controlled.

T-junctions are important part of most of buildings. In this kind of
structure, bottleneck flow, merging flow or split flow are all
possible to take place in different situations. Especially around
the corners, pedestrian behavior is much more complex. Pedestrian
flow characteristics in this kind of geometry are significant to
study. However, except for some studies using a Cellular Automaton
model \cite{Tajima2002,Peng2011}, there are few empirical studies
directly considering the fundamental diagram of pedestrian flow in
T-junctions.

In this study, series of well-controlled laboratory experiments are
carried out to investigate the pedestrian movement in T-junctions.
The Voronoi method is accepted to analyze the fundamental diagram
and field profiles of density, velocity and specific flow.

\section{Experiment setup and trajectory extraction}

\begin{figure}[t]
  \centerline{
  \includegraphics[width=\linewidth]{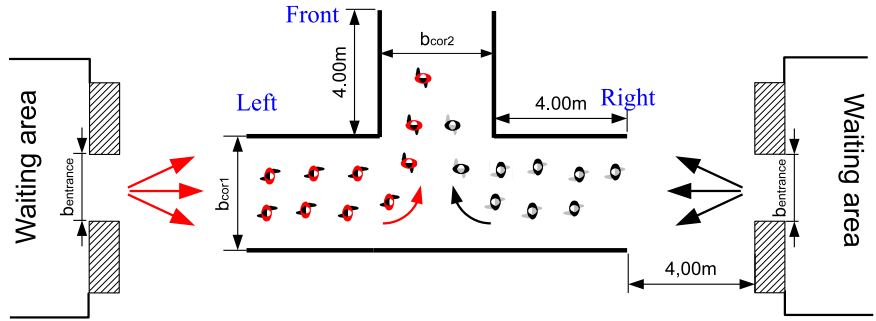}
  }
  \caption{Setup of T-junction experiments.}
  \label{fig:setup}
\end{figure}

Seven runs of experiments were carried out in a T-junction with
corridor width $b_{\rm cor1}= b_{\rm cor2}= 2.4~m$.
Fig.~\ref{fig:setup} shows the sketch of the experiment setup.
Pedestrian streams move from two branches oppositely and then merge
into the main stream at the T-junction. To regulate the pedestrian
density in the corridor, the width of the entrance $b_{\rm
entrance}$ was changed in each run from $0.5~m$ to $2.4~m$, see
details in Table~\ref{table1t}. For simplicity, the left and right
entrances have the same width. In this way, we guarantee the
symmetry of the two branch streams. More than 300 pedestrians
participated in the experiments, which makes the duration of the run
long enough to obtain stationary states. The average age and body
height of the tested persons was $25 \pm 5.7$ years and $1.76 \pm
0.09~m$ (range from $1.49~m$ to $2.01~m$), respectively. They mostly
consisted of German students of both genders. The free velocity $v_0
= 1.55 \pm 0.18~m/s$ was obtained by measuring 42 participants' free
movement.

\begin{table}[htbp]
 \centering\caption{\label{table1t} Parameters for the T-junction experiments }
 \begin{tabular}{cccccc}
  \toprule
  Runs & Name & $b_{\rm cor1}$ [$m$] & $b_{\rm entrance}$ [$m$] & $b_{\rm cor2}$ [$m$] & $N_l+N_r$ \\
  \midrule
  1 & $T$-240-050-240 & 2.40 & 0.50 & 2.40 & 67 + 67 \\
  2 & $T$-240-060-240 & 2.40 & 0.60 & 2.40 & 66 + 66 \\
  3 & $T$-240-080-240 & 2.40 & 0.80 & 2.40 & 114 + 114 \\
  4 & $T$-240-100-240 & 2.40 & 1.00 & 2.40 & 104 + 104 \\
  5 & $T$-240-120-240 & 2.40 & 1.20 & 2.40 & 152 + 153 \\
  6 & $T$-240-150-240 & 2.40 & 1.50 & 2.40 & 153 + 152 \\
  7 & $T$-240-240-240 & 2.40 & 2.40 & 2.40 & 151 + 152 \\
  \bottomrule
 \end{tabular}
\end{table}

At the beginning, the participants were held within waiting areas in
the left and right side. When the experiment starts, they pass
through a $4~m$ passage into the corridor simultaneously and merge
into the main stream at the vertical corner. The passage was used as
a buffer to minimize the effect of the entrance. In this way, the
pedestrian flow in the corridor was nearly homogeneous over its
entire width. When a pedestrian leaves the main corridor, he or she
returned to the waiting area for the next run.

\begin{figure}[t]
\centerline{
  \subfigure[Snapshot]{\includegraphics[width=0.34\linewidth]{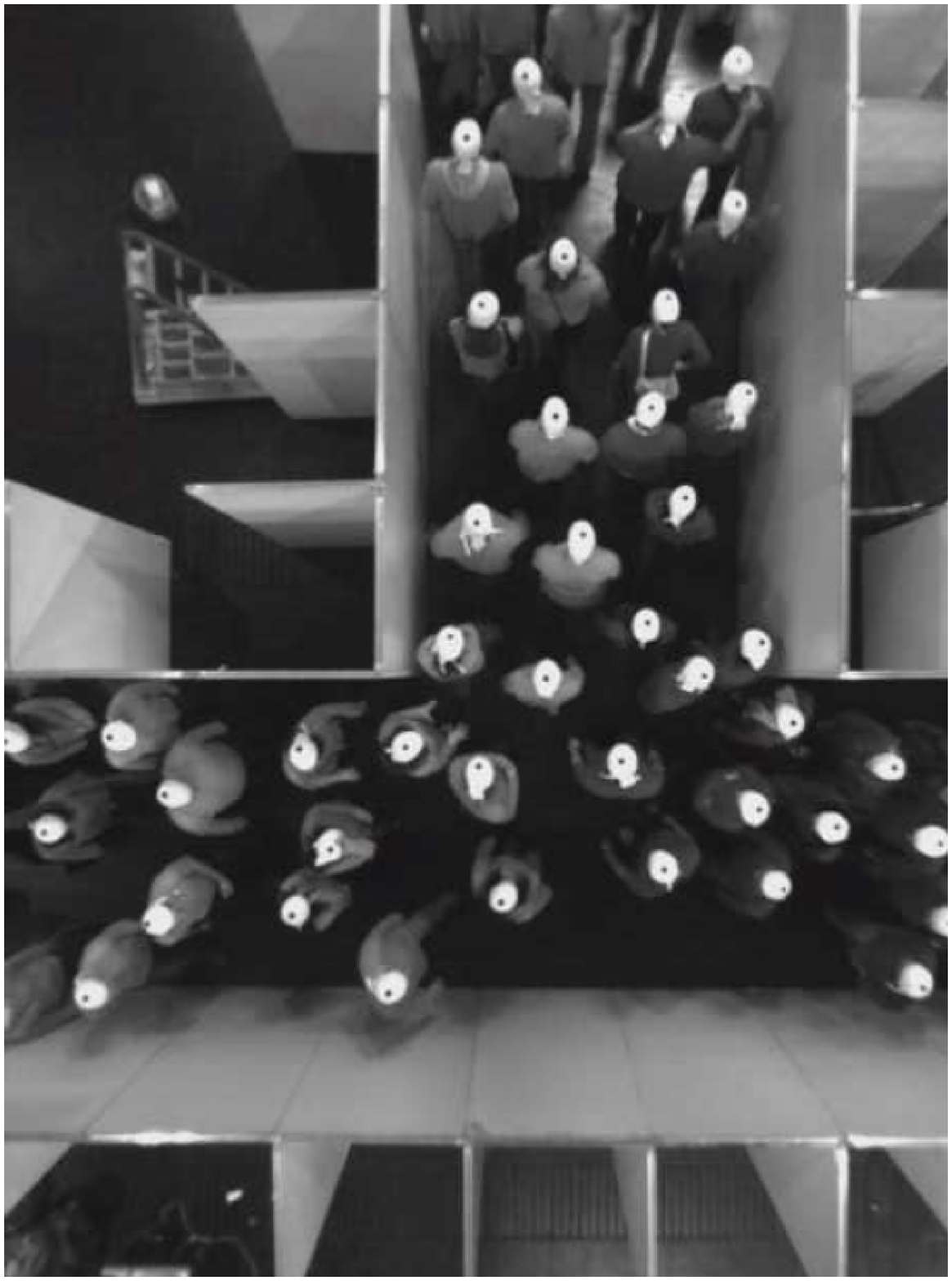}}
  \subfigure[Pedestrian trajectories]{\includegraphics[width=0.66\linewidth]{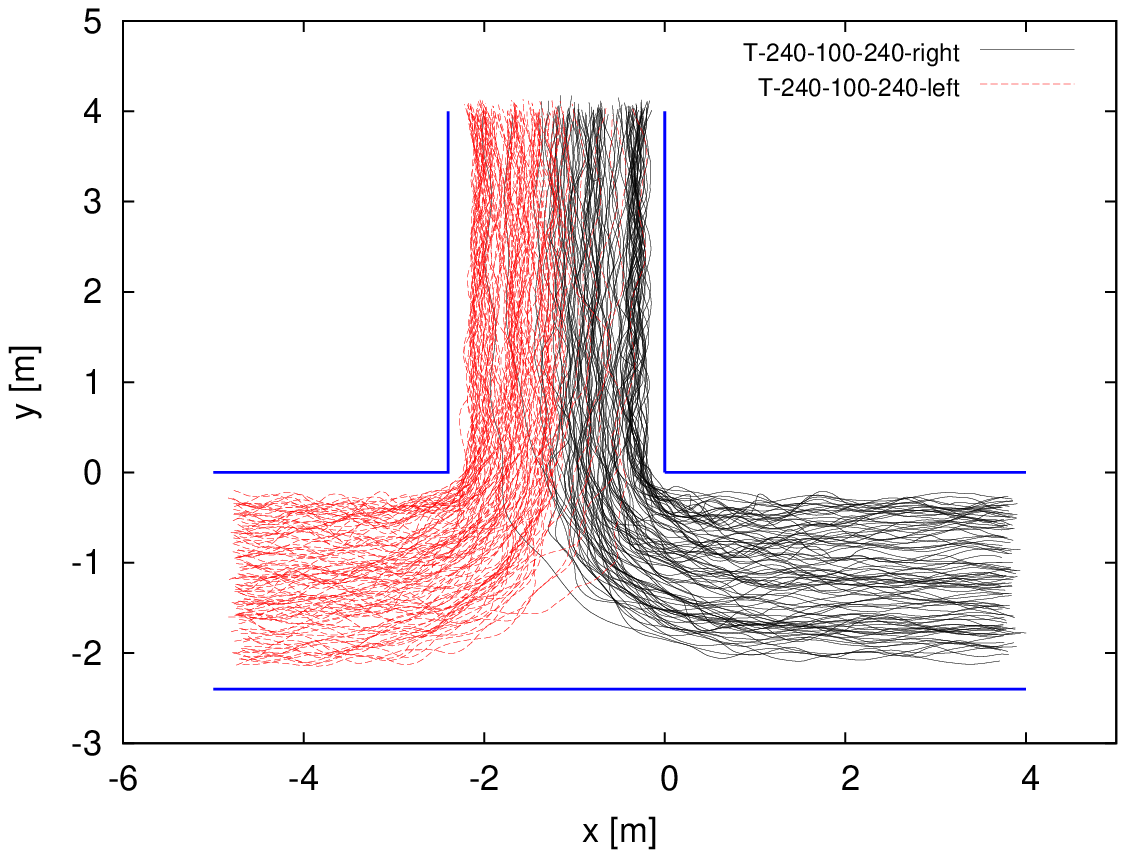}}
}
   \caption{Trajectories and snapshot from T-junction experiment.}
  \label{fig:bsp}
\end{figure}

The whole processes of the experiments were recorded by two
synchronized stereo cameras of type Bumblebee XB3 (manufactured by
Point Grey). They were mounted on the rack of the ceiling $7.84~m$
above the floor with the viewing direction perpendicular to the
floor. The cameras have a resolution of 1280 $\times$ 960 pixels and
a frame rate of 16 $fps$ (corresponding to 0.0625 second per frame.
To cover the complete region, the left and the right part of the
scenario were recorded by the two cameras separately (see
Fig.~\ref{fig:bsp}(a)). The overlapping field of view of the stereo
system is $\alpha = 64^{\circ}$ at the average head distance of
about $6~m$ from the cameras. With the above-mentioned height range,
all pedestrians can be seen without occlusion at any time.

To make a highly precise analysis, accurate pedestrian trajectories
were automatically extracted from video recordings using the
software {\em PeTrack} \cite{Boltes2010}. Lens distortion and
perspective view are taken into account in this program.
Fig.~\ref{fig:bsp}(b) shows the pedestrian trajectories for one run
of the experiments. From these trajectories, pedestrian
characteristics including flow, density, velocity and individual
distances at any time and position can be determined.

\section{Experiment analysis}

In previous studies, different measurement methods were used to
limit the comparability and fluctuation of the data. E.G. Helbing et
al. proposed a Gaussian, distance-dependent weight function
\cite{Helbing2007} to measure the local density and local velocity.
Predtechenskii and Milinskii \cite{Predtechenskii1978} used a
dimensionless definition to consider different body sizes and Fruin
introduced the "Pedestrian Area Module" \cite{Fruin1971}. This study
focuses on the Voronoi method proposed in
\cite{Steffen2010a,Zhanga}, where the density distribution can be
assigned to each pedestrian. This method permits examination on
scales smaller than the pedestrians for its high spatial resolution.

\subsection{Measurement methodology}

At a given time $t$, the Voronoi diagram can be generated from the
positions of each pedestrian. It contains a set of Voronoi cells for
each pedestrian $i$. The cell area, $A_i$, can be thought as the
personal space belonging to each pedestrian $i$. Then, the density
and velocity distribution over space can be defined as
\begin{equation}\label{eq1}
\rho_{xy} = 1/A_i \quad and \quad v_{xy}={v_i(t)}\qquad \mbox{if
$(x,y)\in A_i$}\,
\end{equation}
where $v_i(t)$ is the instantaneous velocity of each person (see
\cite{Zhanga}). The Voronoi density and velocity for the measurement
area $A_m$ is defined as
\begin{equation}\label{eq2}
\langle \rho \rangle_v(x,y,t)=\frac{\iint{\rho_{xy}dxdy}}{A_m}\, ,
\end{equation}
\begin{equation}\label{eq3}
\langle v \rangle_v(x,y,t)=\frac{\iint{v_{xy}dxdy}}{A_m}\, .
\end{equation}
The specific flow
\begin{equation}\label{eq4}
J_s(x,y,t)=\langle \rho \rangle_v(x,y,t) \cdot \langle v
\rangle_v(x,y,t)
\end{equation}
can also be calculated using the Voronoi density and velocity.

\subsection{Results and analysis}

To analyze the spatial dependence of density, velocity and specific
flow precisely, we use the Voronoi method to measure these
quantities in areas smaller than the size of pedestrians. We
calculate the Voronoi density, velocity and specific flow over small
regions ($10~cm \times 10~cm$) each frame. Then the spatiotemporal
profiles of density ($\overline{\rho}(x,y)$), velocity
($\overline{v}(x,y)$) and specific flow ($\overline{J_s}(x,y)$) can
be obtained over the stationary state separately for each run as
follows:

\begin{equation}\label{eq5}
\overline{\rho}(x,y) = \frac{\int_{t_1}^{t_2} \langle \rho
\rangle_v(x,y,t)dt}{t_2-t_1} ,
\end{equation}

\begin{equation}\label{eq6}
\overline{v}(x,y) = \frac{\int_{t_1}^{t_2} \langle v
\rangle_v(x,y,t)dt}{t_2-t_1} ,
\end{equation}

\begin{equation}\label{eq5}
\overline{J_s}(x,y) = \overline{\rho}(x,y) \cdot \overline{v}(x,y) .
\end{equation}

\begin{figure}
\centering\subfigure[Density profile]{
\includegraphics[width=0.45\linewidth]{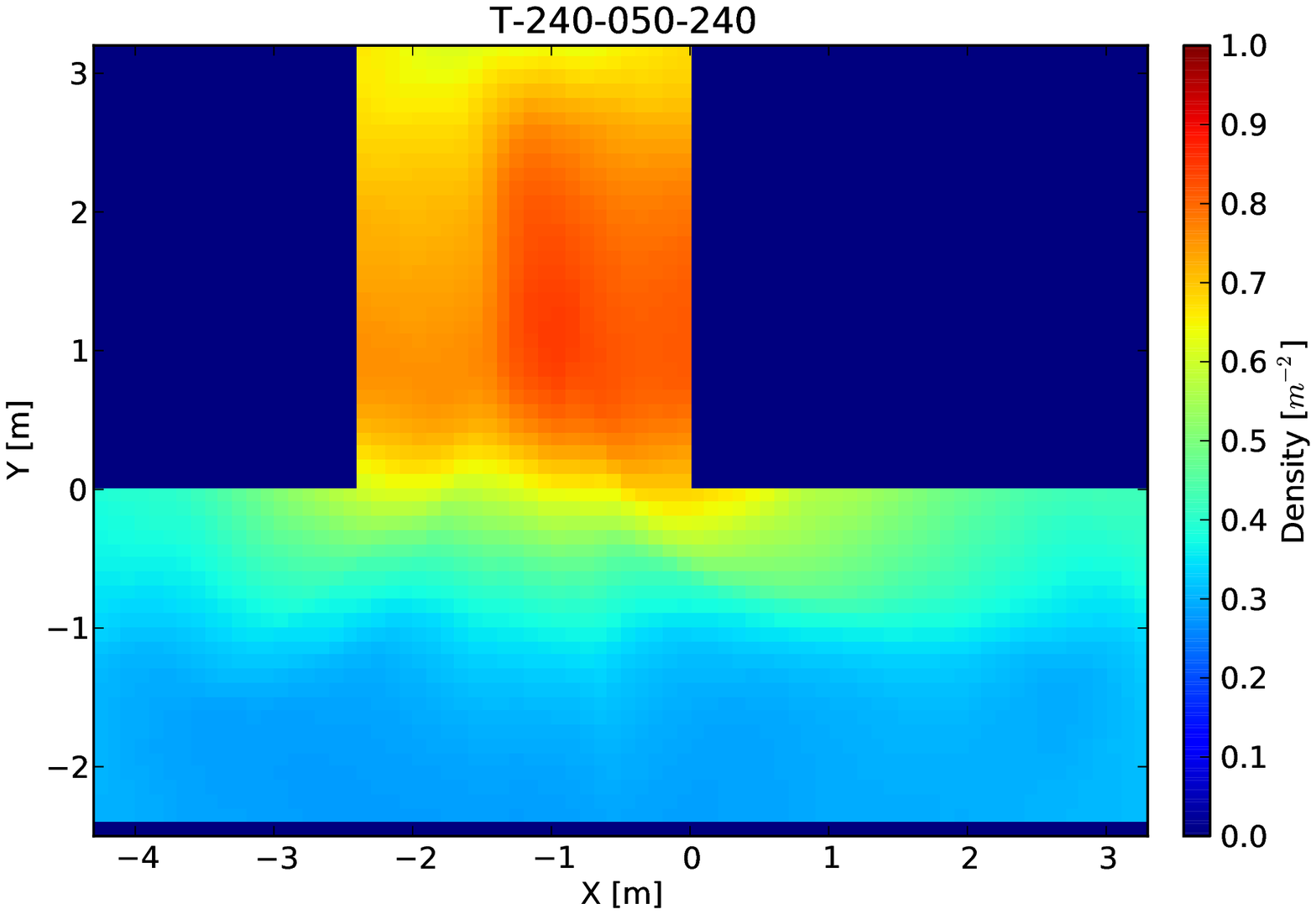}
\includegraphics[width=0.45\linewidth]{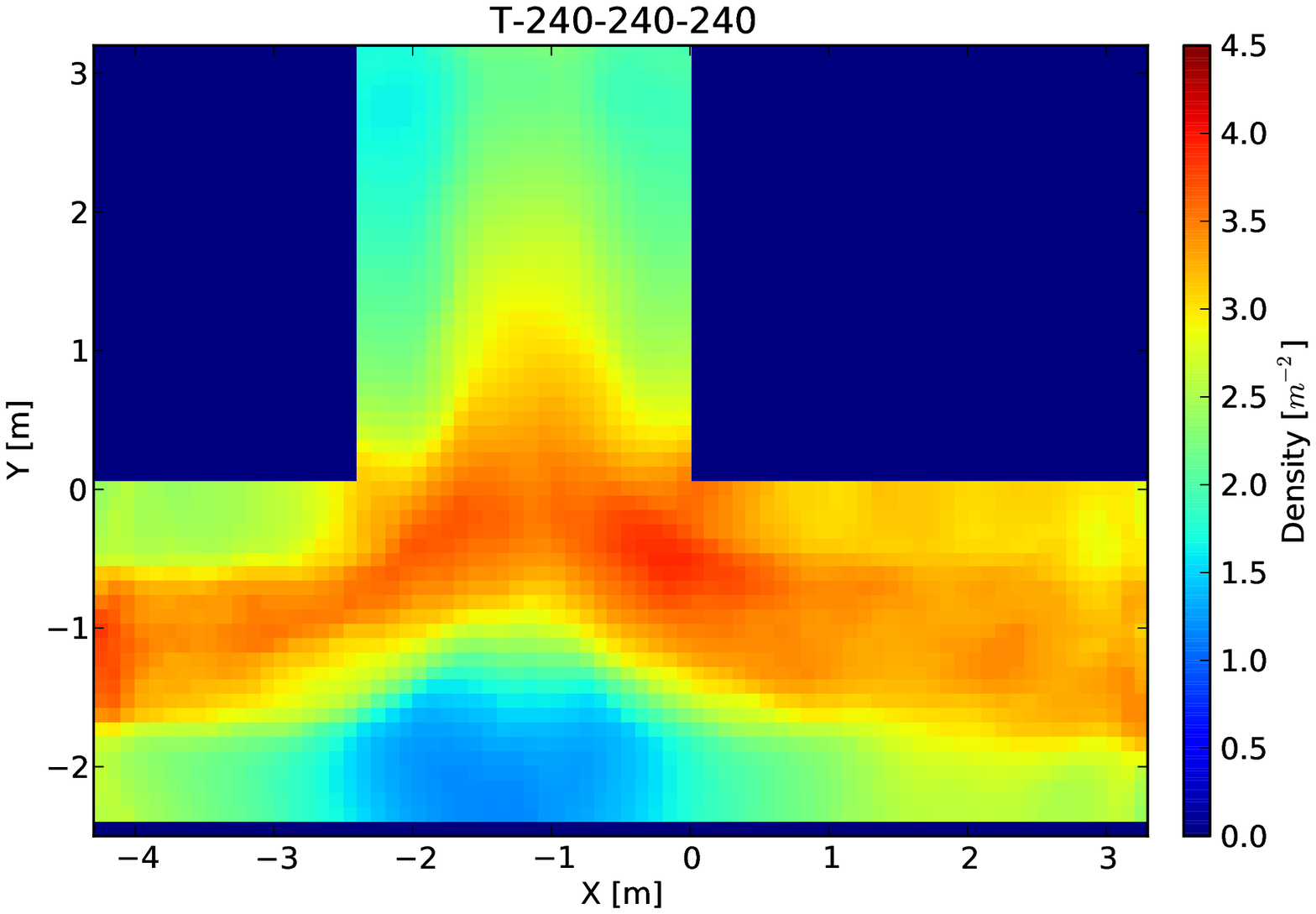}
} \centering\subfigure[Velocity profile]{
\includegraphics[width=0.45\linewidth]{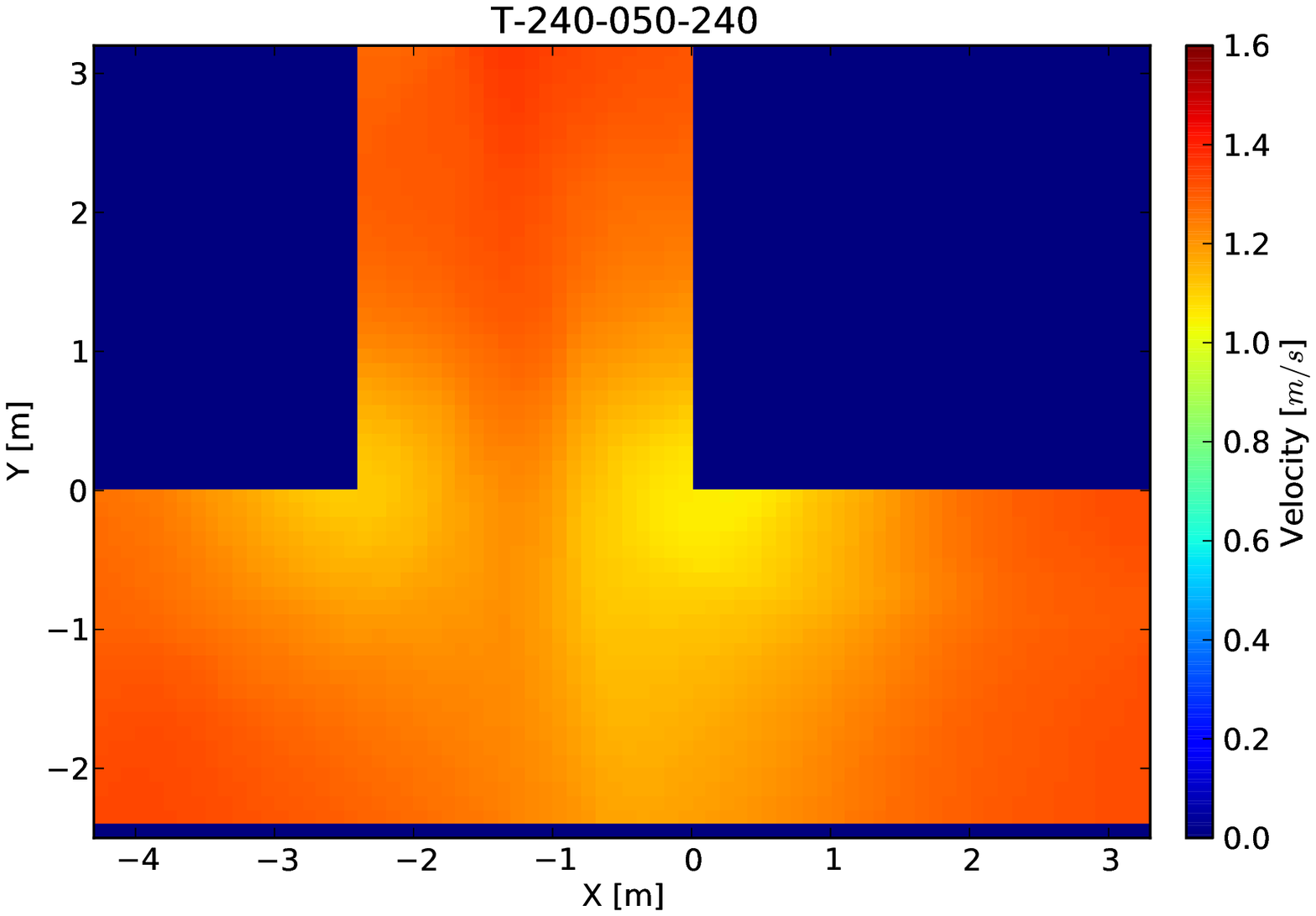}
\includegraphics[width=0.45\linewidth]{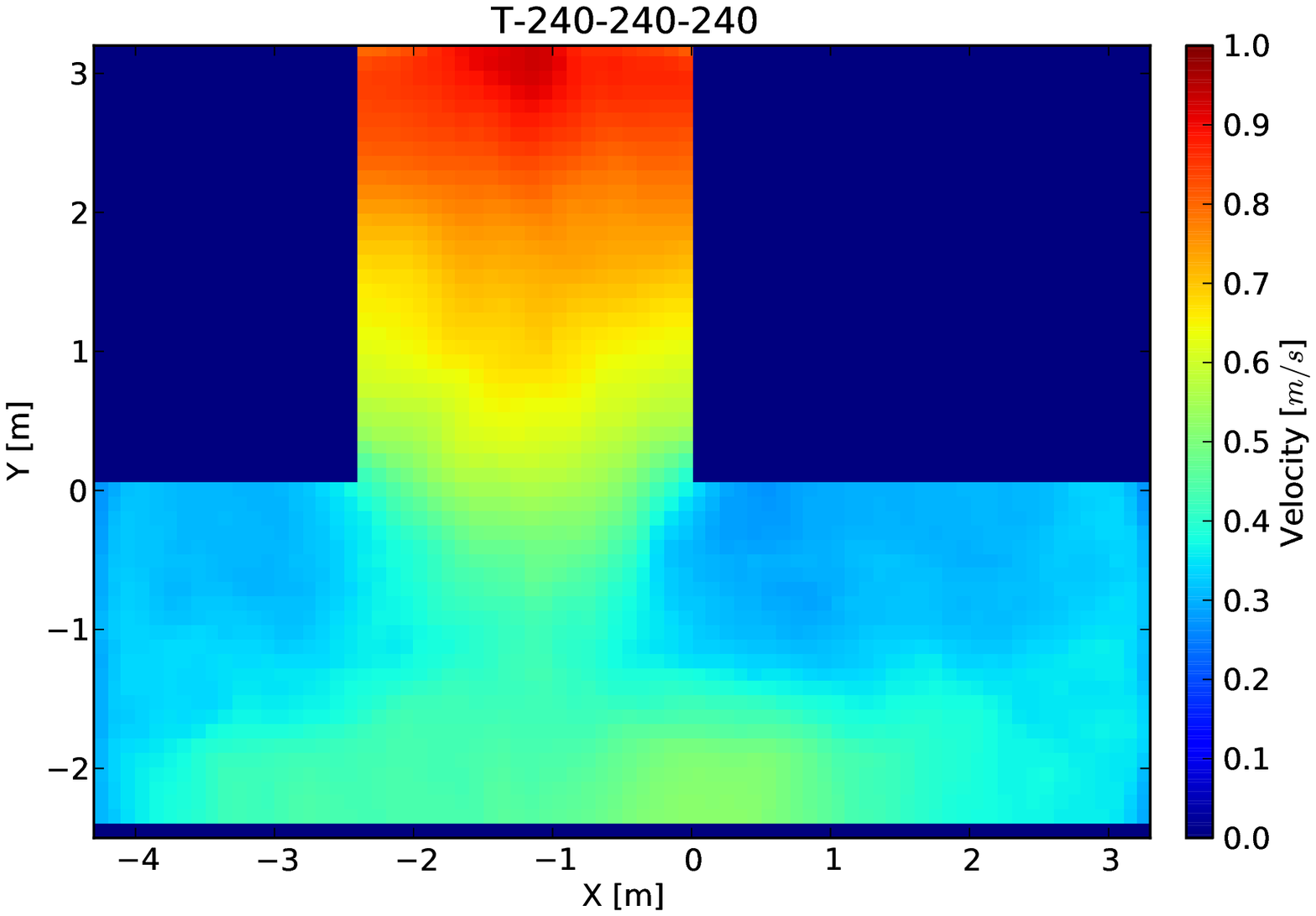}}
\centering\subfigure[Specific flow profile]{
\includegraphics[width=0.45\linewidth]{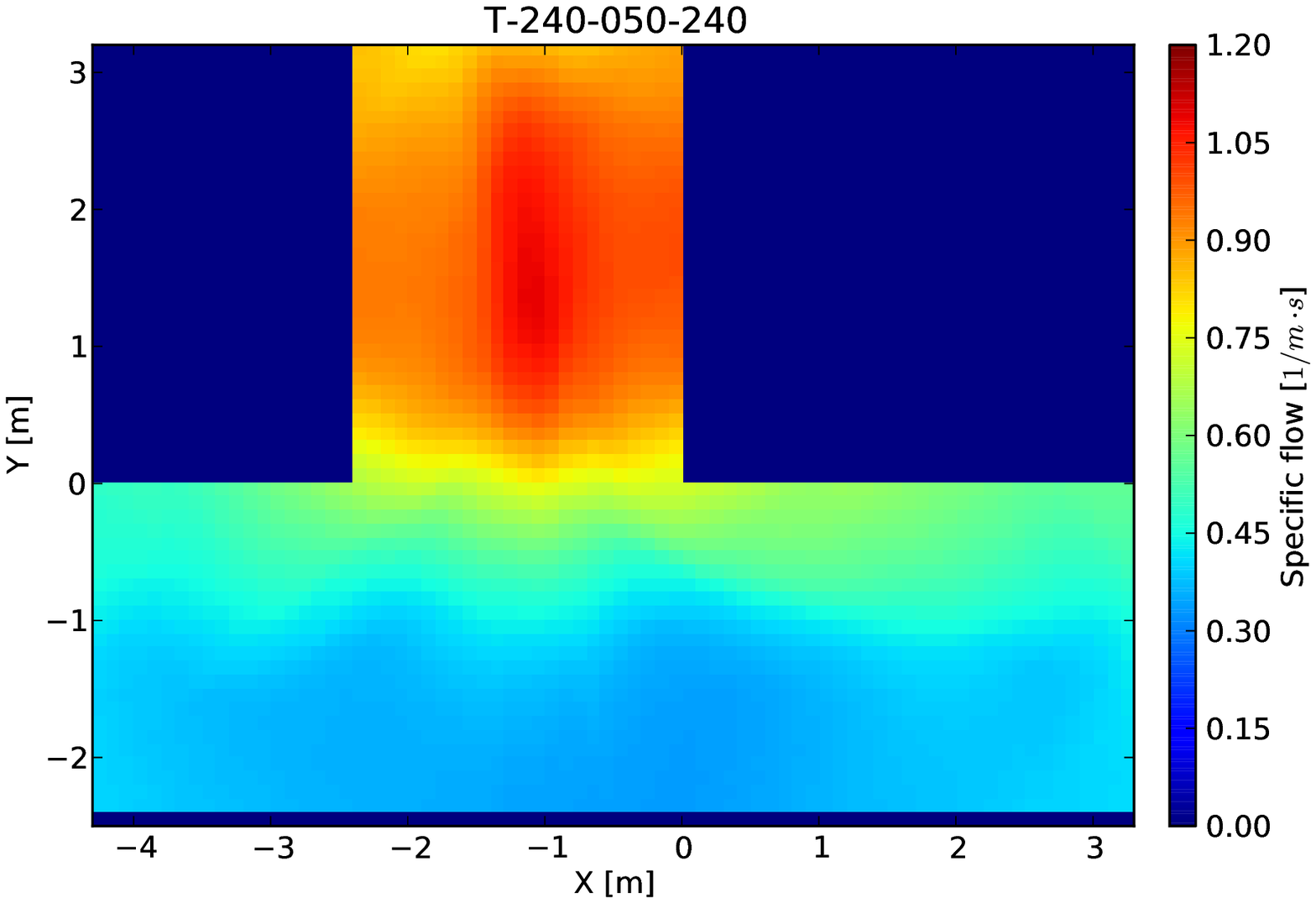}
\includegraphics[width=0.45\linewidth]{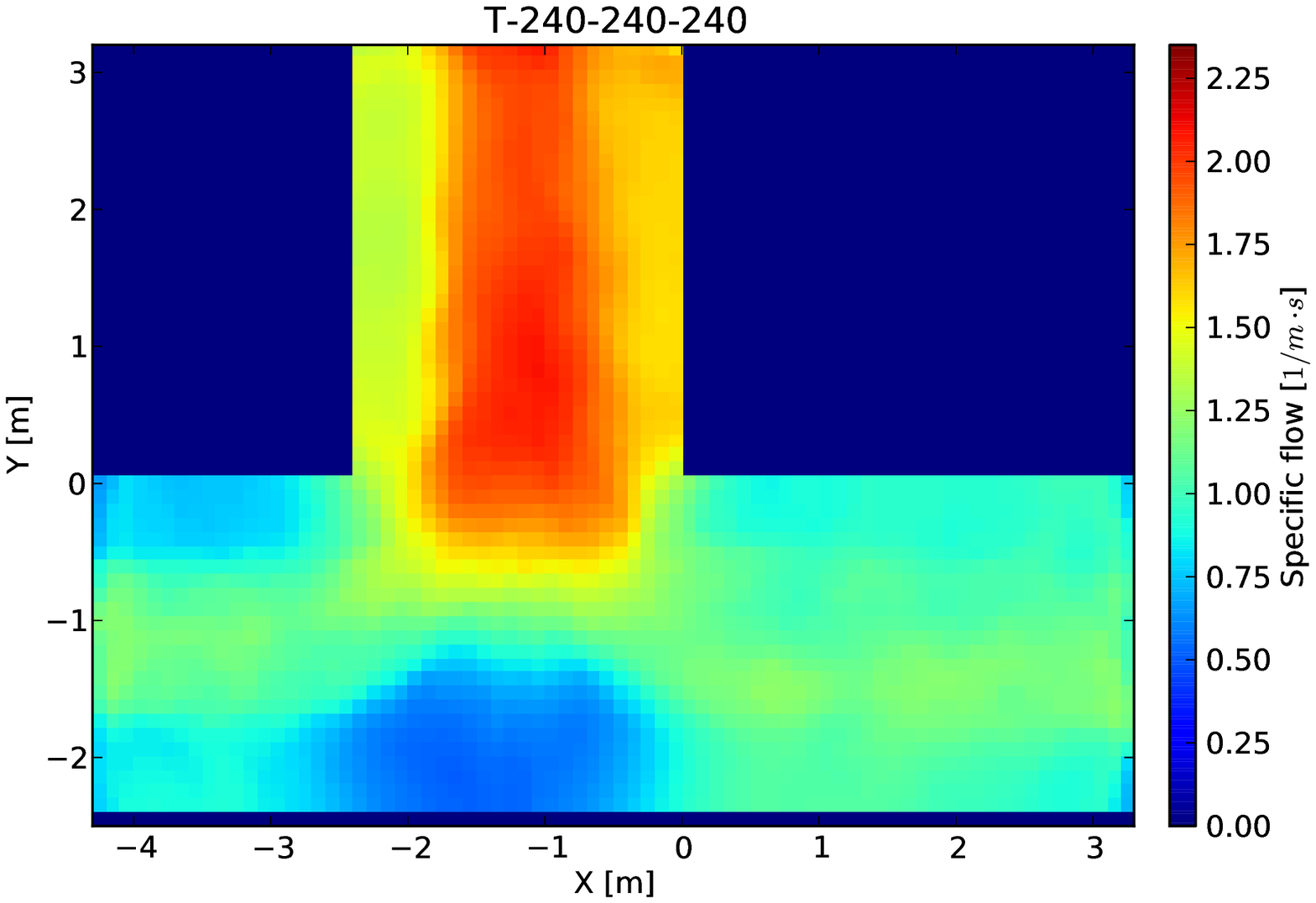}}
\caption{\label{profile}The profiles of density, velocity and
specific flow in T-junction obtained from Voronoi method. The
resolution of the profiles is $10~cm\times10~cm$.}
\end{figure}

Fig.~\ref{profile} shows the profiles for two runs of the
experiments under the situations of low density ($T$-240-050-240)
and high density ($T$-240-240-240), respectively. These profiles
provide new insights into the spatiotemporal dynamics of the motion
and the sensitivity of the quantities to other potential factors.
The density distribution in T-junction is not homogeneous both for
low and high density situations. For the former, the higher density
region locates at the main stream after merging. Whereas for the
latter, the higher density region appears near the junction and the
lowest density region locates at a small triangle area, where the
left and right branches begin to merge. For both of the two
situations, the densities in the branches are not uniform and are
higher over the inner side, especially near the corners. In other
words, pedestrians prefer to move along the shorter and smoother
path. Moreover the density profile shows obvious boundary effects at
high density situation. The spatiotemporal variation of the velocity
is different. At low density situation, the velocity profile is
nearly homogeneous all over the T-junction except the places near
corner. Pedestrians move in free velocity and slow down round the
corner. While for high density condition, the velocity distribution
is not uniform any more. The velocities of main stream is higher
obviously than that in the branches. Boundary effect does not occur
and the velocities after merging increase along the movement
direction persistently. By comparison of the specific flow profiles
at the two different situations, the highest flow regions are both
observed at the center of the main stream after the merging. The
different is that the region extends further into the area where the
two branches start to merge for the high density case, but for the
other case it doesn't. This indicates that the merging process in
front of the exit corridor leads to a flow restriction. Causes for
the restriction of the flow must be located outside the region of
highest flow.

These profiles demonstrate that density and velocity measurements
are sensitive to the size and location of the measurement area. For
the comparison of measurements (e.g. for model validation or
calibration), it is necessary to specify precisely the size and
position of the measurement area.

\begin{figure}
\centering\includegraphics[width=0.48\linewidth]{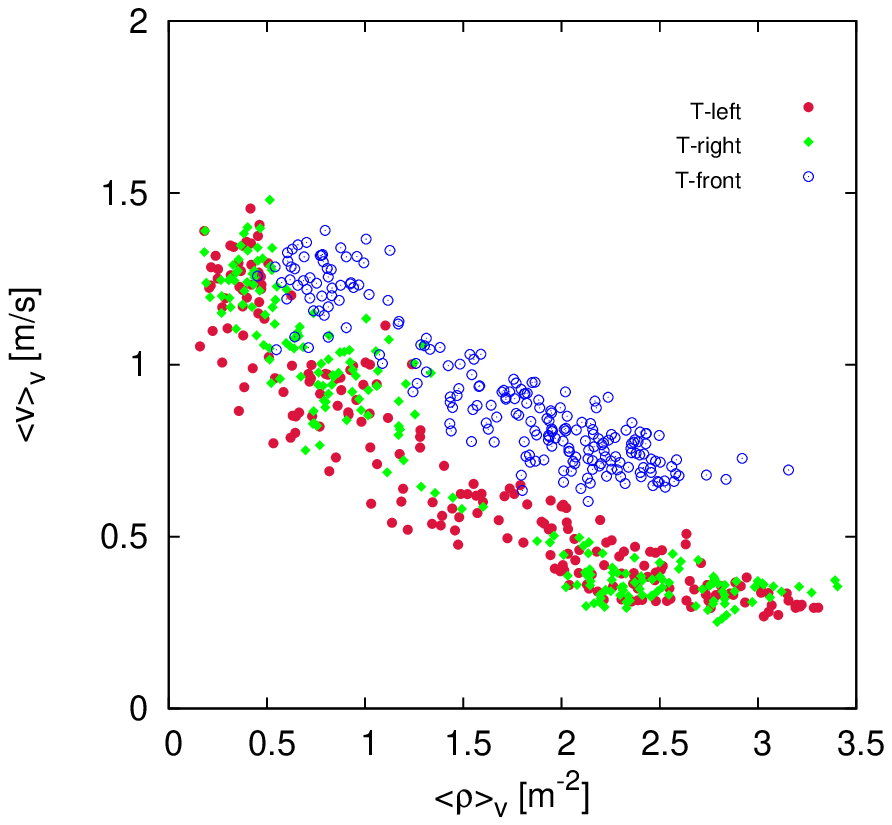}
\centering\includegraphics[width=0.48\linewidth]{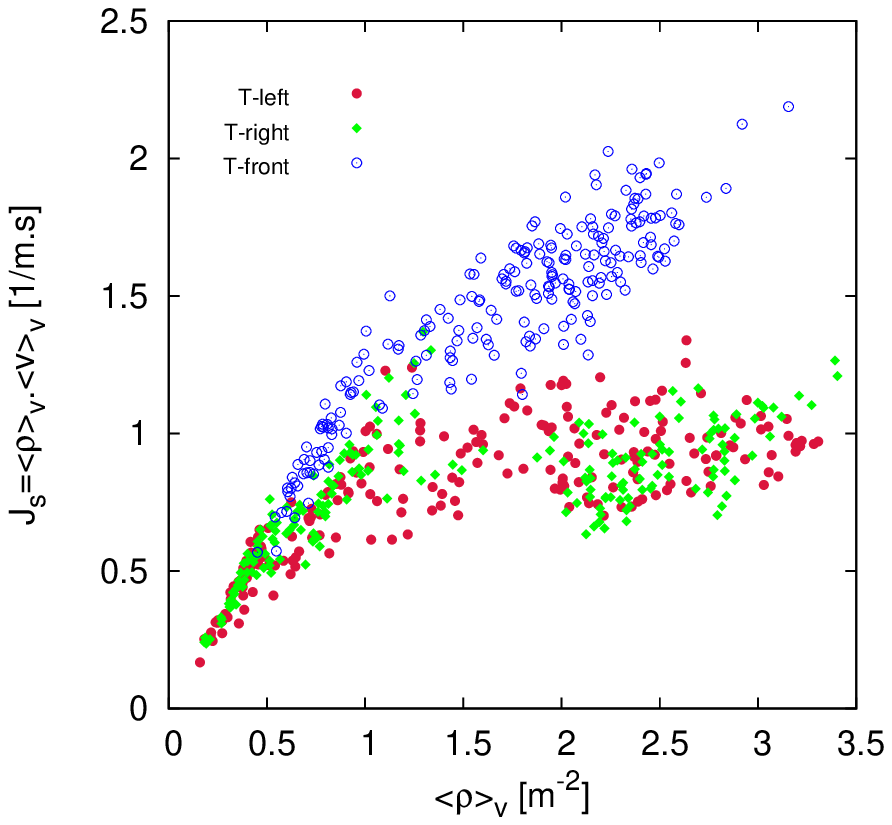}
\caption{\label{FD} Fundamental diagram of pedestrian flow at
different measurement areas in T-junction.}
\end{figure}

In Fig.~\ref{FD}, we compare the fundamental diagrams obtained from
all T-junction experiments. The data assigned with 'T-left' and
'T-right' are measured in the areas before the streams merge, while
the data assigned with 'T front' are measured in the region where
the streams have already merged. The locations of these measurement
areas can be seen in Table.~\ref{tab:mesurarea}. For ease of
comparison, we choose these measurement areas with the same size
($4.8~m^{-2}$). One finds that the fundamental diagrams of the two
branches match well. However, for densities $\rho > 0.5~{m}^{-2}$
the velocities in the 'right' and 'left' part of the T-junction
(T-left and T-right) are significantly lower than the velocities
measured after the merging of the streams (T-front). This
discrepancy becomes more distinct in the relation between density
and specific flow. In the main stream (T-front), the specific flow
increases with the density $\rho$ till $2.5~m^{-2}$. While in the
branches, the specific flow nearly remains constant for density
$\rho$ between $1.5~m^{-2}$ and $3.5~m^{-2}$. Thus, there seems no
unique fundamental diagram which describes the relation between
velocity and density for the complete system.

For this difference, we can only offer assumptions regarding the
causes. One is based on behavior of pedestrians. Congestion occurs
at the end of the branches, where the region of maximum density
appears. Pedestrians stand in a jam in front of the merging and
could not perceive where the congestion disperse or whether the jam
lasts after the merging. In such situation, it is questionable
whether an urge or a push will lead to a benefit. Thus an optimal
usage of the available space becomes unimportant. Otherwise, the
situation totally changes if the location of dissolution becomes
apparent. Then a certain urge or an optimal usage of the available
space makes sense and could lead to a benefit. They will move in a
relatively active way. That's maybe the reason why the velocities
after merging are higher than that in front of merging at the same
density. Whether this explanation is plausible could be answered by
a comparison of these data with experimental data at a corner
without the merging. This comparison is in preparation.

\begin{table}[t]
  \caption{The location of measurement area in T-junction}
  \centerline{
  \begin{tabular}{cc}   \hline\noalign{\smallskip}
  \textbf{Measurement area}\ & \textbf{Range [$m$]}\\\hline\noalign{\smallskip}
  T-left & $x\in[-4.5,-2.5],~y\in[-2.4,0]$ \\
  T-right  & $x\in[1.0,3.0],~y\in[-2.4,0]$ \\
  T-front & $x\in[-2.4,0],~y\in[1.0,3.0]$ \\\hline\noalign{\medskip}
  \end{tabular}
} \label{tab:mesurarea}
\end{table}

%
%

%
%



\printindex
\end{document}